\documentclass[twocolumn,showpacs,preprintnumbers,pre,superscriptaddress]{revtex4-1}
\usepackage{graphicx}
\usepackage{amssymb}
\usepackage{amsmath}
\usepackage{amsfonts,color}
\usepackage{caption}
\usepackage{mathrsfs}
\usepackage{graphicx}

\usepackage{dcolumn}
\usepackage{bm}
\usepackage{subfig}
\usepackage{color}
\usepackage{verbatim}

\newcommand{\ket}[1]{\left|#1\right\rangle}
\newcommand{\overlap}[2]{\left\langle #1|#2\right\rangle}

\newcommand{\norm}[1]{\left| #1 \right|}

\begin{document}


\title{Work Distributions in 1-D Fermions and Bosons with Dual Contact Interactions}

\author{Bin Wang}
\affiliation{Yuanpei College, Peking University, Beijing 100871, China}

\author{Jingning Zhang}
\affiliation{Institute for Interdisciplinary Information Sciences, Tsinghua University, Beijing 100084, China}

\author{H. T. Quan}\email[Email:]{htquan@pku.edu.cn}
\affiliation{School of Physics, Peking University, Beijing 100871, China}
\affiliation{Collaborative Innovation Center of Quantum Matter, Beijing 100871, China}

\date{\today}

\begin{abstract}
We extend the well-known static duality \cite{girardeau1960relationship, cheon1999fermion} between 1-D Bosons and 1-D Fermions to the dynamical version. By utilizing this dynamical duality we find the duality of non-equilibrium work distributions between interacting 1-D bosonic (Lieb-Liniger model) and 1-D fermionic (Cheon-Shigehara model) systems with dual contact interactions. As a special case, the work distribution of the Tonks-Girardeau (TG) gas is identical to that of 1-D free fermionic system even though their momentum distributions are significantly different. In the classical limit, the work distributions of Lieb-Liniger models (Cheon-Shigehara models) with arbitrary coupling strength converge to that of the 1-D noninteracting distinguishable particles, although their elemetary excitations (quasi-particles) obey different statistics, e.g. the Bose-Einstein, the Fermi-Dirac and the fractional statistics. We also present numerical results of the work distributions of Lieb-Liniger model with various coupling strengths, which demonstrate the convergence of work distributions in the classical limit.

\end{abstract}

\maketitle


\section{\label{sec:1}Introduction}
Nonequilibrium phenomena in quantum many-body systems are among the most fundamental and intriguing phenomena in physics. In the past two decades, nonequilibrium work relations \cite{jarzynski2011equalities}, including Jarzynski equality \cite{jarzynski1997nonequilibrium}, Crooks fluctuation theorem \cite{crooks1999entropy}, Hummer-Szabo relation  \cite{hummer2001free} and Hatano-Sasa relation \cite{hatano2001steady}, have attracted lots of attention. These relations, collectively known as fluctuation theorems \cite{seifert2012stochastic}, have shed new light on our understanding of the far-from equilibrium statistical physics. The validity of the classical version of these relations have been tested and confirmed in various systems \cite{trepagnier2004experimental,liphardt2002equilibrium,collin2005verification,bustamante2005nonequilibrium}. In the past few years, the quantum version of these relations have been studied extensively \cite{esposito2009nonequilibrium,campisi2011colloquium}. It is found that the quantum work determined by the so-called two-time energy measurement scenario, one at the beginning and one at the end of the protocol, turns out to be effective when there is no heat transfer between the system and the bath. More recently, some experiments have been carried out to measure the work distributions and verify the fluctuation relations in the quantum regime \cite{batalhao2014experimental, an2014experimental, campisi2013employing, campisi2015nonequilibrium}. These theoretical and experimental studies have opened a new avenue to study the nonequilibrium thermodynamics in the quantum regime \cite{jarzynski2015,poletti2015,gasparinetti2016,liu2016,tapio2016,seifert2016,ueda2016,paraoanu2017,acin2017,talkner2017,gongjiangbin2017,wisniacki2017,dong2017}.

Previous explorations of quantum work relations have been mostly focused on either single-particle systems, such as  the parametric harmonic oscillator \cite{deffner2008nonequilibrium, del2013more}, a single particle in a time-dependent piston \cite{quan2012validity,zhu2016,garcia2018semiclassical} and two-level systems \cite{quan2008microscopic}, or a sudden quench process of a quantum many-body system \cite{dora2012,silva2012,silva2013b}. Although work distributions in ideal quantum gases have been considered in Ref. \cite{gong2014interference}, few efforts have been devoted to the study of work distributions of interacting quantum many-body systems in an arbitrary driven process (but see Refs. \cite{yi2011work,yi2012work,silva2013,wang2017understanding,einax2009work}). However in  real world, most quantum systems are interacting many-body systems and the driven processes are usually not sudden quench processes. The quantum correlation and interaction make it difficult to study the non-equilibrium dynamics, because quantum scattering of identical particles display both single and many-body interference \cite{tichy2014interference}. The theoretical study of the work distributions of interacting quantum many-body systems in arbitrary driven processes, challenging but realistic, would be very helpful for improving our understanding about the effects of quantum statistics and interactions on quantum work.

Although it is difficult to study interacting quantum many-body systems in three dimensions, fantastic results have been obtained in many-body quantum systems in reduced dimensions \cite{eisert2015quantum, langen2013local}. One example is the bosonoic gas with $\delta$ interaction \cite{slshanii2001,olshanii2003short,de2014solution}, which was first proposed by Lieb and Liniger \cite{lieb1963exact,lieb1963exact2}. Its limiting case, known as Tonk-Girardeau (TG) gas \cite{girardeau1960relationship}, has been realized in the optical lattice experiment \cite{paredes2004tonks, kinoshita2004observation}. While most of the theoretical studies of this model are focused on the static properties \cite{yang1967some,yang1969,cheon1999fermion, girardeau2006anyon,settino2017signatures}, few are devoted to the studies of time-dependent properties \cite{girardeau2000dark}. On the other hand, more recently, developments in experimental techniques have made it possible to explore such systems with tunable coupling strength \cite{trotzky2012probing, meinert2015probing}, which might enable one to experimentally measure the non-equilibrium work and study nonequilibrium statistical mechanics in the quantum many-body system. Meanwhile, Lieb-Liniger model is one of few exactly solvable quantum many-body models \cite{yu2015understanding}, which provides deep insights to many interesting and important collective features of many-body phenomena, such as quantum integrability. Hence, Lieb-Liniger model also serves as an insightful example to illustrate the nonequlibrium statistical properties of interacting quantum many-body systems.

In this article, by extending the previous static duality between TG gas and the free Fermions to the dynamical version, we study the relation of work distributions between 1-D bosonic (Lieb-Liniger model) and 1-D fermionic (Cheon-Shigehara model) systems with dual contact interactions. We find that the Bose-Fermi duality and the duality in interactions ``cancel" each other. As a result, the work distributions for the two systems are identical even though their momentum distributions are significantly different. In addition we find that in the classical limit, the work distributions of the Lieb-Liniger model (Cheon-Shigehara model) converge to that of noninteracting distinguishable particles, irrespective of the coupling strength of the Lieb-Liniger model (Cheon-Shigehara model). This article is organized as follows. In Sec. II, we introduce the models and compare the moemntum distributions of the two models. In Sec. III, we derive the dynamical duality between 1D bosonic and fermionic gases with dual contact interactions, as an extension of previous work on hard-core Bosons \cite{girardeau1960relationship, cheon1999fermion}. In Sec. IV, we prove the duality of work distributions between 1-D Bosons and 1-D Fermions with dual contact interactions. In Sec. V, we find that in the classical limit, quantum work distributions of Lieb-Liniger model converge to that of the noninteracting distinguishable particles, irrespective of the coupling strength $C$. In section VI, we discuss our main results and conclude the paper. Note that throughout this paper, we would use superscript ``B'' (``F'') to denote the bosonic (fermionic) system.

\section{1-D Bosonic and Fermionic Systems with dual contact Interactions}

Consider $N$ identical Bosons in one dimension with the contact interaction \cite{lieb1963exact,lieb1963exact2} subjected to a time-dependent external potential $V_{\rm ext}(\cdot, \lambda(t))$, with $\lambda(t)$ being the externally controlled time-dependent work parameter. The system is described by the following Lieb-Liniger Hamiltonian,
\begin{equation}
\label{eqn:bose_hamil}
\hat{H}^B(\lambda(t))=\sum_{i=1}^{N}\frac{\hat{p_i}^2}{2m}+\sum_{i=1}^N V_{\rm ext}(\hat x_i,\lambda(t))+\sum_{i<j}\delta(\hat x_i-\hat x_j,C),
\end{equation}
where $\hat x_i$ and $\hat{p_i}$ are the position and momentum operators for the $i$-th particle. The two-body contact interaction is denoted as $\delta(x,C)\equiv C\delta(x)$, where $\delta(\cdot)$ is the Dirac delta function and the constant $C$ characterizes the coupling strength. Note that work is applied to the system when the work parameter $\lambda(t)$  is varied.

The corresponding 1-D fermionic system consists of $N$ identical fermions subjected to the same time-dependent external potential. The two-body interaction in the fermionic system that corresponds to the bosnonic contact interaction is the generalized point-like potential \cite{cheon1999fermion}, which can be simplified in the short-range limit \footnote{The limit about $a$ is subtle, to be more rigorous, please see \cite{cheon1998realizing}} as follows,
\begin{eqnarray*}
\varepsilon(x,\frac{1}{C})= \lim_{a\rightarrow 0^+}\left(\frac{C}{2}-\frac{1}{a}\right)[\delta(x-a)+\delta(x+a)],
\end{eqnarray*}
when it is applied to antisymmetric fermonic wave functions. Note that this is a sensible renormalized zero-range limit which allows the discontinuity in the wave-function \cite{cheon1998realizing}.

The many-body Hamiltonian of the above fermionc system reads
\begin{equation}
\label{eqn:fermi_hamil}
\hat{H}^F(\lambda(t))=\sum_{i=1}^{N}\frac{\hat{p_i}^2}{2m}+\sum_{i=1}^N{V}_{ext}(\hat x_i,\lambda(t))+\sum_{i<j}\varepsilon(\hat x_i-\hat x_j,\frac{1}{C}),
\end{equation}
where $C$ is the coupling strength of the corresponding bosonic system.  It's worth noting that this interaction is attractive when the corresponding bosonic interaction is repulsive $C>0$. Note that we have set the coupling strength in Eqs. (\ref{eqn:bose_hamil}) and (\ref{eqn:fermi_hamil}) to be the inverse of each other.

For an arbitrary fixed work parameter $\lambda$, the time-independent Schr$\ddot{\rm o}$dinger equation of the bosonic (fermionic) systems can be written as follows,
\begin{eqnarray}
\label{eqn:time_independent_schr}
\hat H^{B, F}(\lambda)\phi_n^{B, F}({\mathbf x};\lambda) = E^{B, F}_n(\lambda)\phi_n^{B, F}({\mathbf x}; \lambda),
\end{eqnarray}
where $\phi_n^{B, F}({\mathbf x};\lambda)$ and $E^{B, F}_n(\lambda)$ denote the eigenstates and eigenenergies of the bosonic (fermionic) system with ${\mathbf x}\equiv\left(x_1,x_2,\ldots,x_N\right)$. As shown in Ref. \cite{cheon1999fermion}, the bosonic and fermionic systems with dual contact interaction have the same eigenenergies, $E^B_n(\lambda)=E_n^F(\lambda)\equiv E_n(\lambda)$, and their eigenstates are related by the following relation,
\begin{eqnarray}
\phi_n^F({\mathbf x}; \lambda)=A({\mathbf x})\phi_n^B({\mathbf x}; \lambda),
\label{duality}
\end{eqnarray}
where $A({\mathbf x})\equiv\prod_{i<j}{\rm sign}(x_j-x_i)$.

\begin{figure*}[htp]
\centering
\includegraphics[width = \textwidth]{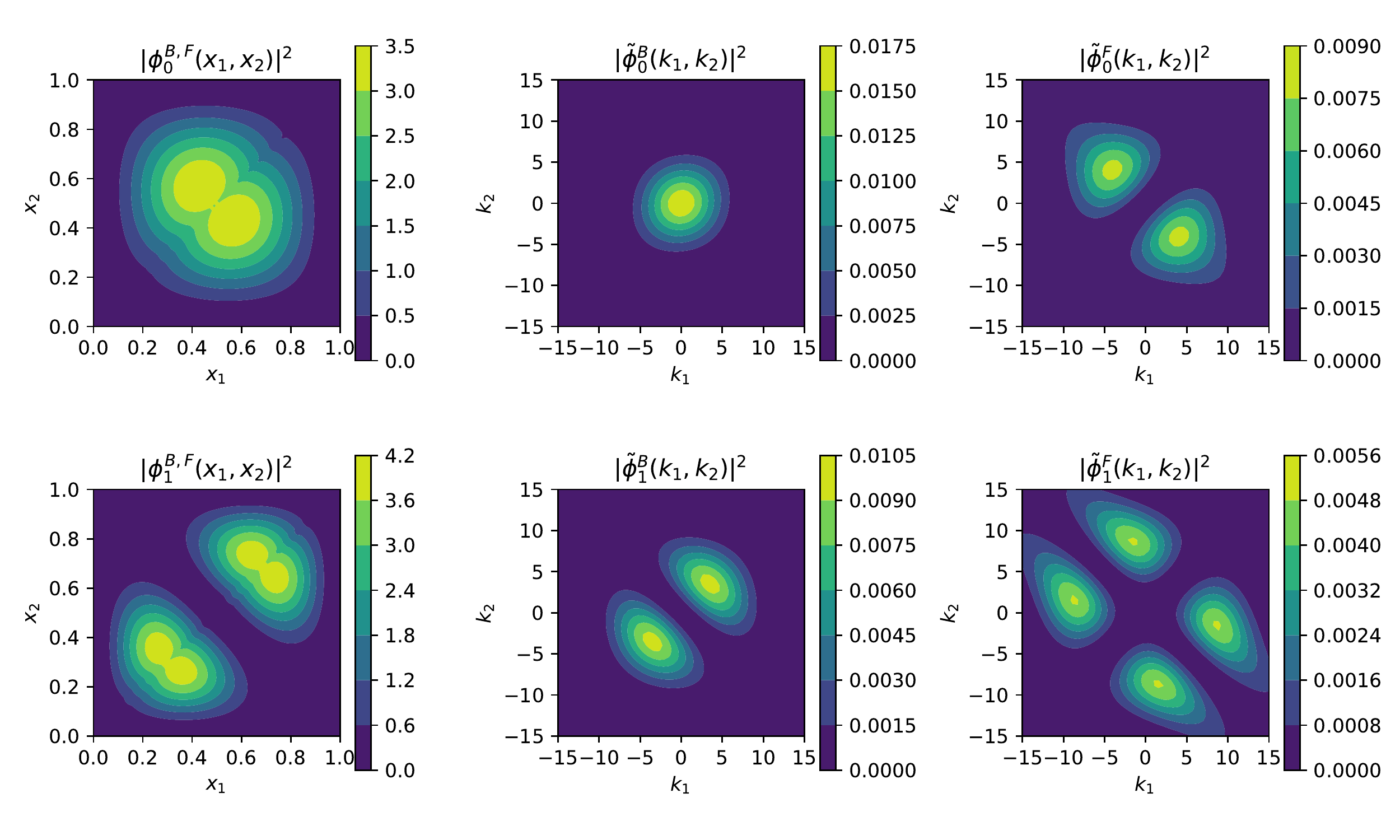}
\caption{The spatial and momentum probability distributions for the ground and the first excited states of two-particle bosonic \eqref{eqn:bose_hamil} or fermionic \eqref{eqn:fermi_hamil} systems confined in a 1-D box. The positions $x_{1, 2}$ and the wave vectors $k_{1, 2}$ are in units of $\lambda$ and $\lambda^{-1}$, respectively, with $\lambda$ being the width of the 1-D box. The dimensionless coupling strength is chosen to be $\alpha\equiv\hbar^{-2}m\lambda C=5$.}
\label{fig:density_distributions}
\end{figure*}

From the mapping of the eigenstates (\ref{duality}), it is straightforward to see that the spatial probability distributions $\norm{\phi_n^{B, F}({\mathbf x};\lambda)}^2$ are the same for the bosonic and the fermionic systems with dual contact interactions. However, due to the existence of $A({\mathbf x})$, the momentum probability distributions $\norm{\tilde\phi_n^{B, F}({\mathbf k};\lambda)}^2$ are quite different, where the wave function in momentum space is obtained by the Fourier transform,
\begin{eqnarray}
\tilde\phi_n^{B, F}({\mathbf k};\lambda)=(2\pi)^{-N/2}\int e^{-i{\mathbf k}\cdot{\mathbf x}}\phi_n^{B, F}({\mathbf x};\lambda)d^N x.
\end{eqnarray}
It has been known for many decades \cite{lenard1964,olshanii1998atomic} that when $C \to +\infty$, the momentum distributions of the corresponding eigenstates of 1-D free Fermions and TG gas will be different. This result is also true for any finite $C$. In Fig.~\ref{fig:density_distributions}, for a finite $C$ we show representative results of the spatial and momentum probability distributions of two-particle bosonic and fermionic systems with dual contact interactions confined in a 1-D box. It can be seen that the eigenstates for Bosons and Fermions have different momentum distributions. Also, we would like to emphasize that the mapping (\ref{duality}) of the eigenstate wavefunctions exists in the position representation only. In the momentum representation there is no such a mapping.

Since the work applied to the system, when the boundary of the box is moved, depends on the momentum of the particles inside the box, naively, one might expect that the work distributions will be different for 1-D bosonic \eqref{eqn:bose_hamil} and fermionic \eqref{eqn:fermi_hamil} systems with dual contact interactions. However, this intuition turns out to be incorrect.

\section{\label{sec:2}dynamical duality of 1D bosons and fermions with dual contact interactions}

In this section we extend the duality about the energy eigenstates \cite{cheon1999fermion} of the two systems to the time evolution of the wave function. Let the initial states of the bosonic and fermionic systems be related by the following relation at $t=0$,
\begin{equation}
\label{eqn:3}
\psi_0^B({\mathbf x})=A({\mathbf x})\psi_0^F({\mathbf x}),
\end{equation}
where $\psi_0^{F(B)}\left({\mathbf x}\right)$ is an arbitrary $N$-body antisymmetric (symmetric) wave function, and $A\left({\mathbf x}\right)\equiv\prod_{i<j}{\rm sign}(x_i-x_j)$ bridges different symmetries between the bosonic and fermionic systems. Here we emphasize that $\psi_0^{B,F}({\mathbf x})$ is not necessarily the energy eigenstate, even though in Ref. \cite{cheon1999fermion} a relation between the energy eigenstates of Eq.  (\ref{eqn:bose_hamil}) and Eq. (\ref{eqn:fermi_hamil}) is established.

The time evolution of the bonsonic and ferminonic systems are governed by the time-dependent Schr$\ddot{\rm o}$dinger equations,
\begin{eqnarray}
\label{eqn:schrodinger}
i\hbar\frac{\partial}{\partial t}\psi^{B, F}({\mathbf x}, t)=\hat H^{B, F}({\mathbf x}, \lambda(t))\psi^{B, F}({\mathbf x}, t),
\end{eqnarray}
where $\psi^{B, F}({\mathbf x}, t)$ is the time-dependent $N$-body wave functions of the bosonic and fermionic systems with initial conditions $\psi^{B,F}({\mathbf x}, 0)=\psi_0^{B, F}({\mathbf x})$. $\hat H^{B, F}({\mathbf x}, \lambda(t))$ are the Hamilonians in Eq. (\ref{eqn:bose_hamil}) and Eq. (\ref{eqn:fermi_hamil}), respectively.

Let us consider a pair of models described by Eqs. (\ref{eqn:bose_hamil}) and (\ref{eqn:fermi_hamil}) respectively. We put both systems in the same 1-D box potential whose boundary moves according to the same predetermined protocol $\lambda(t)$ \footnote{It does not matter that if the external potential $V_{ext}(\hat{x}_i,\lambda(t))$ is not a 1-D box. All the proof goes similar}. The coordinates of the two boundaries of the 1-D box are denoted as $0$ and $\lambda(t)$, respectively. Absorbing boundary conditions are assumed for both systems \footnote{We do not use the periodic boundary condition here because it gets intractable when the number $N$ is odd for the fermionic system.}
\begin{eqnarray}
\label{eqn:absorb_bc}
\left.\psi^{B, F}({\mathbf x},t)\right|_{x_i=0\mbox{ or }\lambda(t)}=0,
\end{eqnarray}
for $i=1, 2\ldots N$.

The dynamical duality of 1-D bosonic and fermionic systems can be expressed as follows. If the initial wave functions of the bosonic and fermionic systems are related by Eq.~(\ref{eqn:3}), the time evolutions of the wave functions at an arbitrary time $t>0$ are related by the same $A(\mathbf x)$
\begin{equation}
\label{eqn:dyanmical-duality}
\psi ^B({\mathbf x}, t)=A({\mathbf x})\psi^F({\mathbf x}, t).
\end{equation}
The detailed proof of this dynamical duality is given in Appendix \ref{Appendix: A}.

Using the dynamical duality \eqref{eqn:dyanmical-duality}, we can easily prove that the time evolutions of the expectation values of any physical observable of these two systems are identical. However, since Bosons and Fermions obey different permutation relations, the distributions of some physical observables, e.g., momentum, of the two systems may be different, as indicated in Refs. \cite{lenard1964,olshanii1998atomic} (see Fig. \ref{fig:density_distributions}).

\section{\label{sec:3}duality of work distributions}
Having established the dynamical duality \eqref{eqn:dyanmical-duality}, we will consider the non-equilibrium work distributions of the bosonic \eqref{eqn:bose_hamil} and fermionic \eqref{eqn:fermi_hamil} systems with dual contact interactions in this section. Without heat transfer, the quantum work for a particular realization is determined by the two energy projective measurements right before and after the driving process,
\begin{equation}
\label{eqn:micro_work}
W_{n_i,n_f} = E_{n_f}(\lambda_f)-E_{n_i}(\lambda_i),
\end{equation}
where $E_{n}(\lambda)$ denotes the $n$-th energy eigenvalue of the many-body system with the work parameter equal to $\lambda$ and $\lambda_{i}(\lambda_{f})$ is the inital (final) value of the work parameter. For a predetermined protocol $\lambda(t)$ with the total time duration $\tau$, $\lambda_i=\lambda(0)$ and $\lambda_f=\lambda(\tau)$. $n_i$ ($n_f$) denotes the quantum number of the initial (final) energy eigenstates. Here the work parameter $\lambda(t)$ is the width of the 1-D box potential.

The distribution function of work can be formally written as \cite{talkner2007fluctuation}
\begin{eqnarray}
\label{eqn:work_dist}
P(W)=\sum_{n_i,n_f} P_i(n_i) P(n_f|n_i) \delta(W-W_{n_i,n_f}),
\end{eqnarray}
where $P_i(n)\equiv Z_i^{-1}\exp\left[-\beta E_{n}(\lambda_i)\right]$ is the thermal equilibrium distribution corresponding to the initial Hamiltonian with the partition function $Z_i\equiv\sum_ne^{-\beta E_n(\lambda_i)}$ and the inverse temperature $\beta = \left(k_BT\right)^{-1}$. $P(n_f|n_i)$ is the transition probability from the initial state $n_i$ to the final state $n_f$.
\begin{eqnarray}
\label{eqn:trans_prob}
P(n_f|n_i)&=&\left|\left\langle n_f(\lambda_f)|\hat U(\tau)|n_i(\lambda_i) \right \rangle\right|^2\\
&=&\left|\overlap{ n_f(\lambda_f)}{\psi_{n_i}(\tau)}\right|^2,\nonumber
\end{eqnarray}
where $\left\{\ket{n(\lambda)}\right\}$ is the $n$-th energy eigenstate of the Hamiltonian corresponding to the work parameter $\lambda$, whose energy spectrum is denoted as $\left\{E_n(\lambda)\right\}$. Given the initial state $\ket{n_i(\lambda_i)}$, the final state is $\ket{\psi_{n_i}(\tau)}\equiv\hat U(\tau)\ket{n_i(\lambda_i)}$, where the evolution operator $\hat U(t)$ satisfies $i\hbar\frac{\partial\hat U(t)}{\partial t}=\hat{H}(\lambda(t))\hat U(t)$. The summation in Eq. (\ref{eqn:work_dist}) is over all initial and final energy eigenstates, namely all possible ``trajectories".

As shown in Ref.~\cite{cheon1999fermion}, the energy spectra of the bosonic ($\ref{eqn:bose_hamil}$) and fermionic ($\ref{eqn:fermi_hamil}$) systems with dual contact interactions are identical. Thus the following relations hold true at the initial ($t=0$) and final ($t=\tau$) moments of time,
\begin{equation*}
E^F_{n}(\lambda_{i, f})=E^B_{n}(\lambda_{i,f}).
\end{equation*}
Moreover, the eigenstates are also related by the following relations,
\begin{eqnarray}
&\phi_n^{B}({\mathbf x},\lambda_{i, f})=A({\mathbf x})\phi_n^{F}({\mathbf x}, \lambda_{i, f}),
\end{eqnarray}
$\phi_n^{B,F}({\mathbf x},\lambda_{i, f})\equiv\overlap{\mathbf x}{n^{B, F}(\lambda_{i,f})}$ is the wave function of the $n$th energy eigenstate. Substituting these relations and Eq. (\ref{eqn:dyanmical-duality}) into Eqs. (\ref{eqn:micro_work})--(\ref{eqn:trans_prob}) and noticing the fact that $\left|A\left({\mathbf x}\right)\right|^2=1$, it is straightforward to obtain
\begin{equation}
\label{eqn:work_distrition_duality}
P^B(W)=P^F(W).
\end{equation}
where $P^B(W)$ ($P^F(W)$) are the work distributions of the bosonic ($\ref{eqn:bose_hamil}$) and fermionic systems ($\ref{eqn:fermi_hamil}$). Note that Eq. (\ref{eqn:work_distrition_duality}) is valid for an arbitrary coupling strength $C$, and an arbitrary driving protocol $\lambda(t)$. Two limiting cases of the duality \eqref{eqn:work_distrition_duality} are listed as follows: (I) When $C \rightarrow \infty $, the Hamiltonian (\ref{eqn:bose_hamil}) describes the TG gas \cite{girardeau1960relationship}, while the Hamiltonian (\ref{eqn:fermi_hamil}) describes 1-D free fermions \footnote{Set $C=2/a$. Since $a \rightarrow 0$, $C \rightarrow \infty $.}. (II) When $C \rightarrow 0$, the Hamiltonian (\ref{eqn:bose_hamil}) describes free Bosons, while the Hamiltonian (\ref{eqn:fermi_hamil}) describes fermionic Tonks-Girardeau gas (FTG gas) \cite{girardeau2006bosonization}. In both cases, the work distributions of the bosonic \eqref{eqn:bose_hamil} and fermionic \eqref{eqn:fermi_hamil} systems with dual contact interactions are identical while their momentum distributions are qualitatively different (see Fig. \ref{fig:density_distributions}) \cite{lenard1964,olshanii1998atomic}. This is the first main result of our paper and it is summarized in Table \ref{tab:1}.
\begin{table}
\caption{Duality of work distributions for the bosonic and fermionic systems with dual contact interactions with various coupling strength $C$. For an arbitrary C, the work distributions of Bosons and Fermions in the same row are identical. Notice that when $C>0$, the $\delta$ interaction is repulsive while the $\varepsilon$ interaction is attractive.}
\label{tab:1}
\begin{ruledtabular}
\begin{tabular}{ccc}
Coupling strength $C$ & Bosons & Fermions\\
\hline
$C>0$ & \text{$\delta$ interaction} & \text{$\varepsilon$ interaction}\\
	  & (\text{Lieb-Liniger \cite{lieb1963exact}}) & (\text{Cheon-Shigehara \cite{cheon1999fermion}})\\
\hline
$C\rightarrow +\infty$ & \text{Impenetrable}&\text{Free}\\
					   & \text{(TG gas \cite{girardeau1960relationship})}&(\text{Fermions})  \\
\hline
$C\rightarrow 0$ & \text{Free} & \text{Strong attraction}\\
				 &	(\text{Bosons})		   & (\text{FTG gas \cite{girardeau2006bosonization}})\\
\end{tabular}
\end{ruledtabular}
\end{table}

\section{\label{sec:4}Convergence of work distributions in the classical limit}
\begin{table}[b]
\caption{Classical limit ($\hbar\rightarrow0$ or $T\rightarrow+\infty$) of the duality relations of work distributions (see Table I). Work distributions of all the systems in this table are identical irrespective of the coupling strength $C$. NDP stands for noninteracting distinguishable particles.}
\label{tab:2}
\begin{ruledtabular}
\begin{tabular}{ccc}
Coupling strength $C$ &Bosons&Fermions\\
\hline
$C>0$ & \text{Impenetrable(Tonks \cite{tonks1936complete})} & \text{Penetrable}\\
$C\rightarrow +\infty$ & \text{Impenetrable (Tonks \cite{tonks1936complete})}&\text{NDP}\\
$C\rightarrow 0$ & \text{NDP} & \text{Penetrable}\\
\end{tabular}
\end{ruledtabular}
\end{table}
\begin{figure*}[htp]
\subfloat[]{
\includegraphics[width=0.5\textwidth]{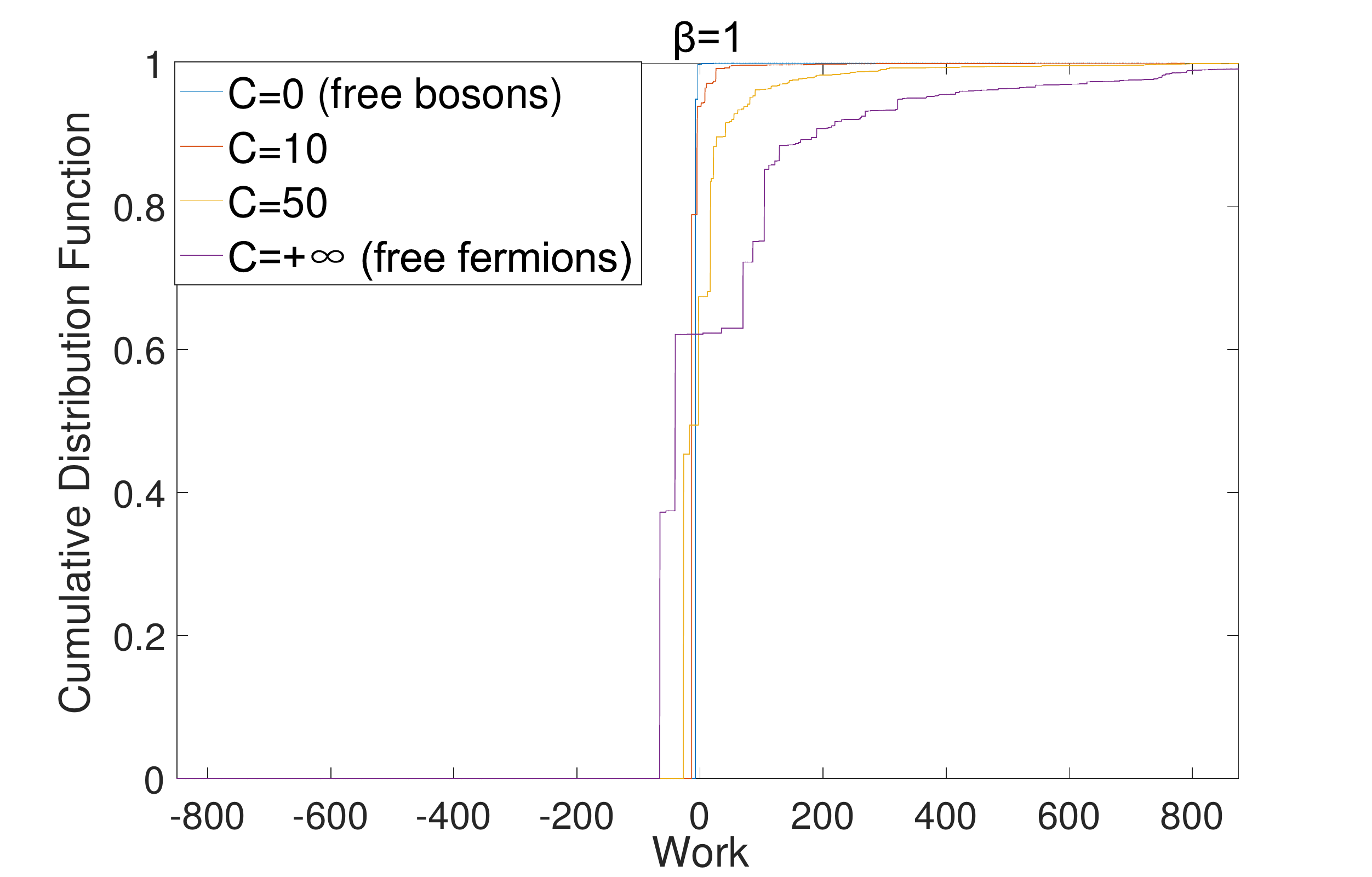}}
\subfloat[]
{\includegraphics[width=0.5\textwidth]{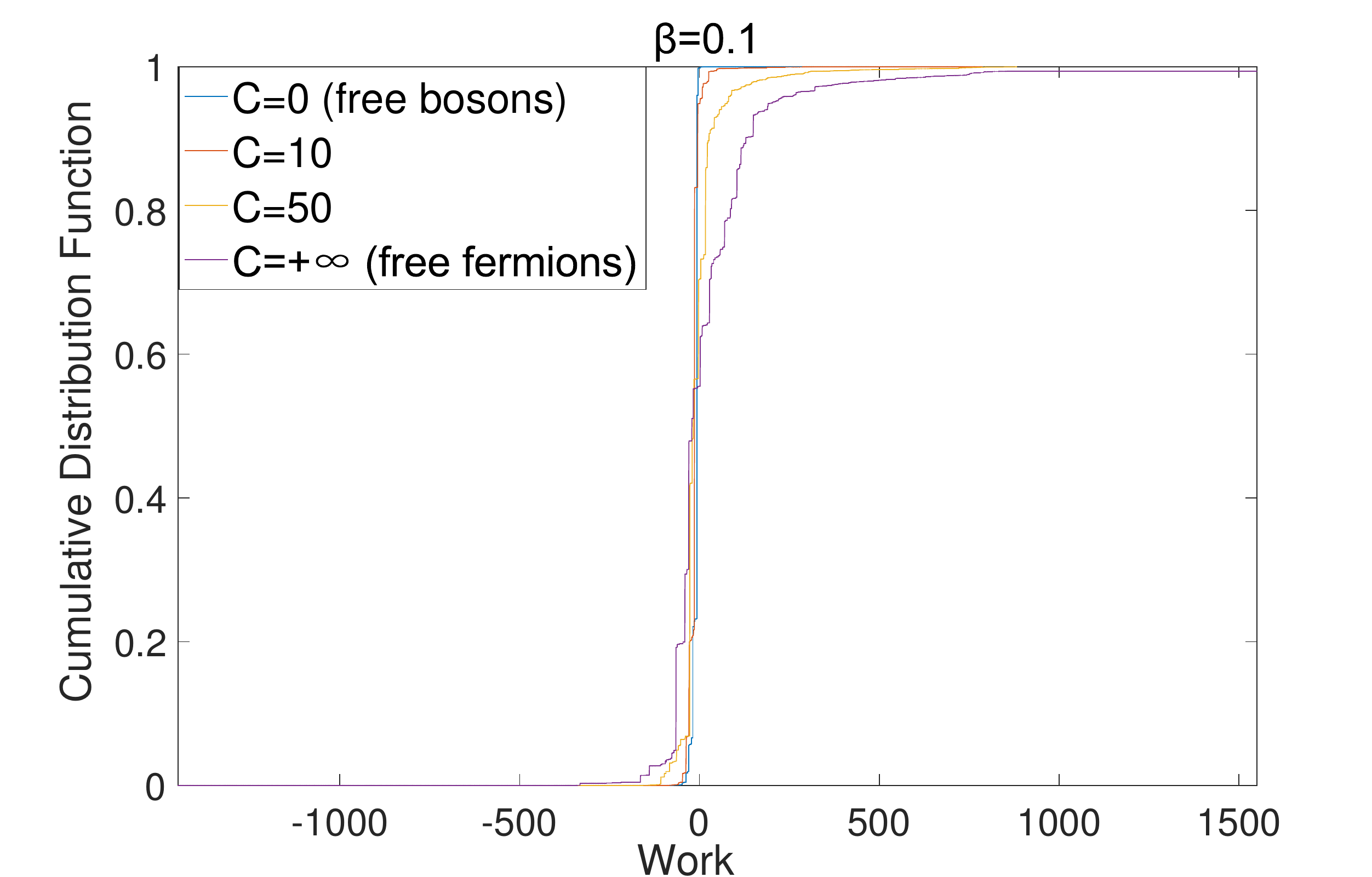}}
\\
\subfloat[]{
\includegraphics[width=0.5\textwidth]{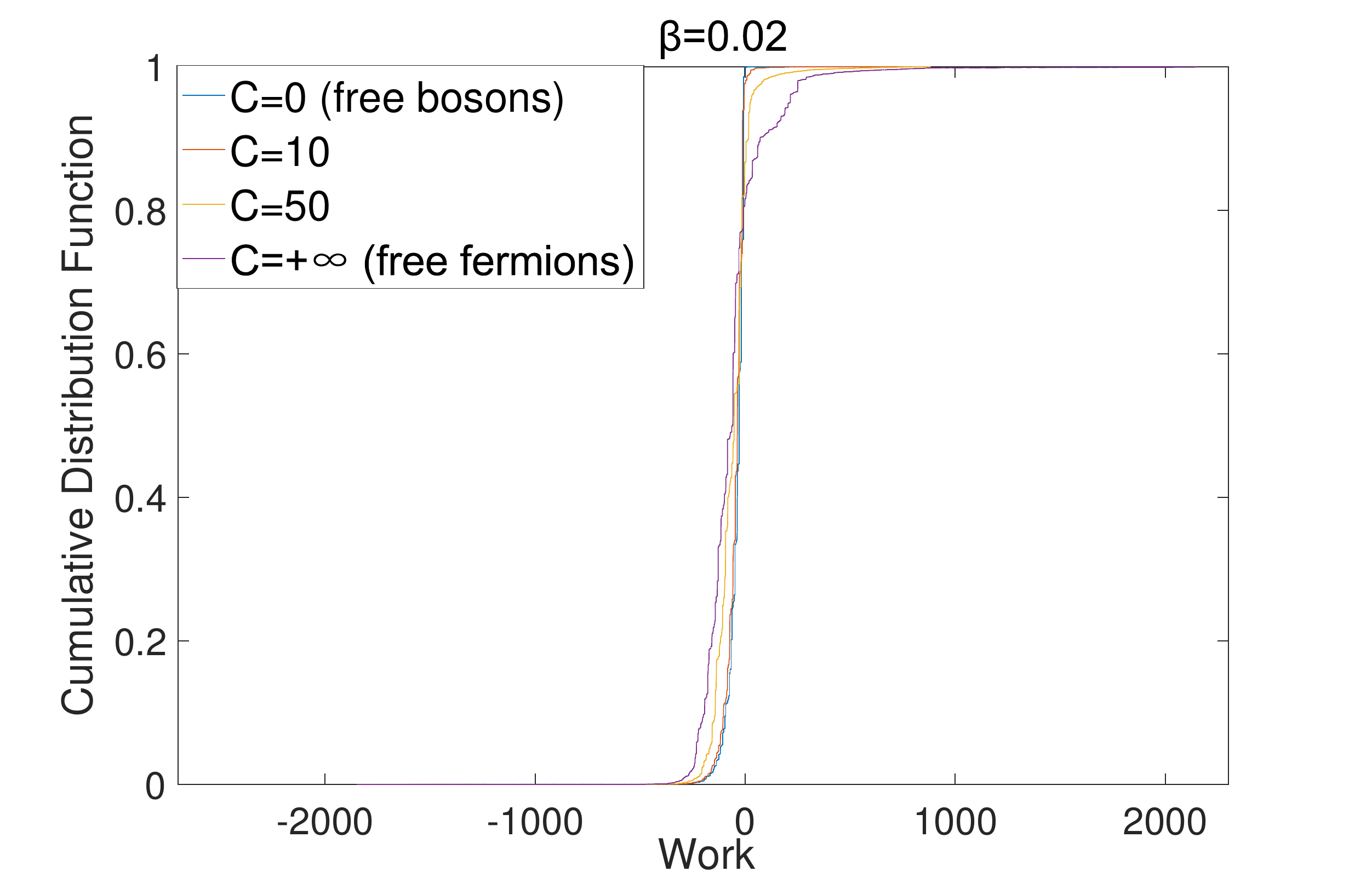}
}
\subfloat[]
{
\includegraphics[width=0.5\textwidth]{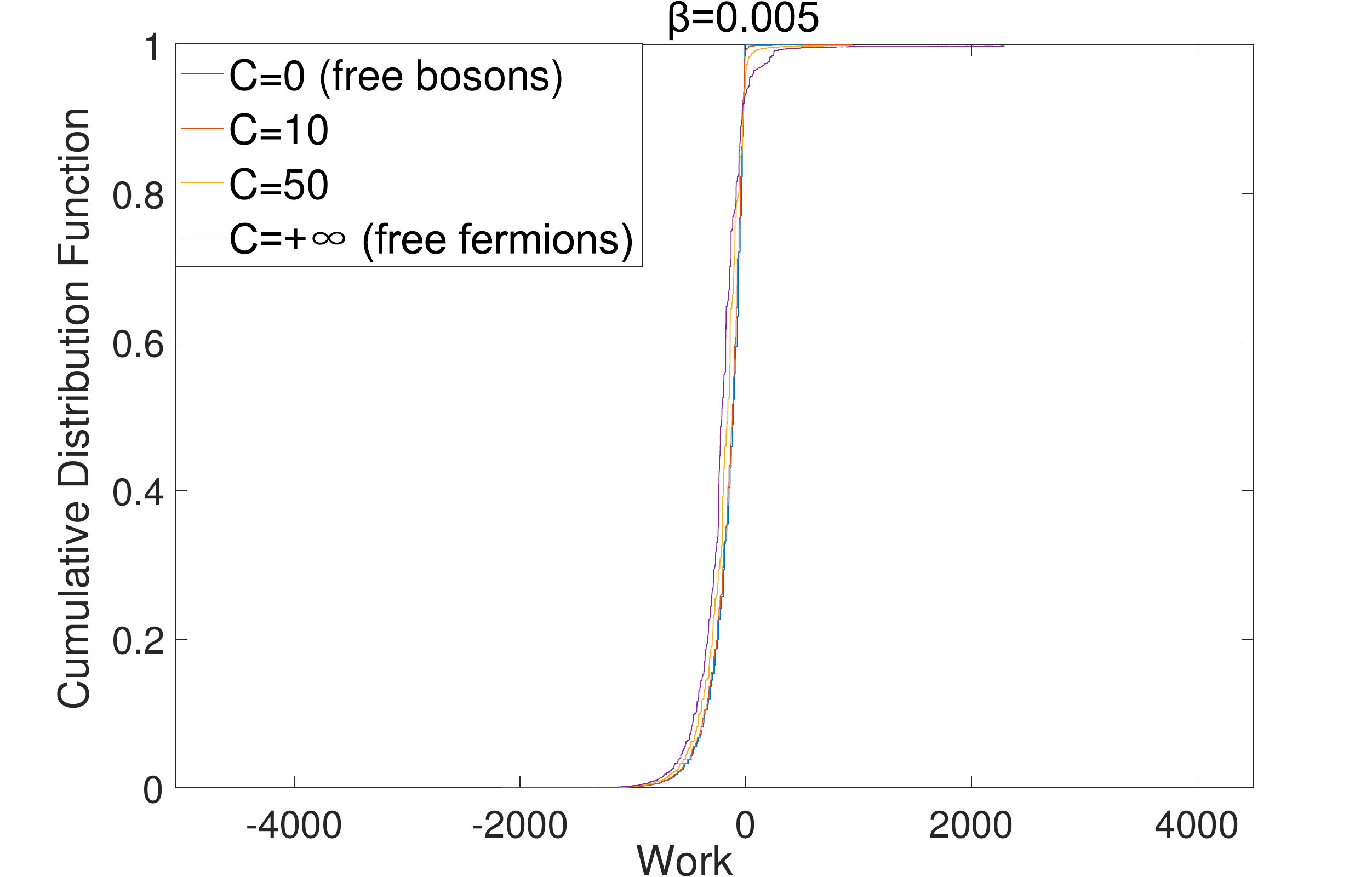}
}
\caption{ \label{fig:2} Work distributions of the Lieb-Liniger models \cite{lieb1963exact,lieb1963exact2} with various coupling strengths $C$ and different initial temperatures. The unit of energy is $\hbar^2/2m\lambda_{0}^2$. The protocol of changing the work parameters is chosen to be $\lambda(t)=1+vt$ with $v=5$, $t\in[0,1]$. The value of the inverse temperature $\beta$ is shown on the top of every plot. All the work distributions converge as the initial temperature goes to infinity.}
\end{figure*}
Having established the duality of the work distributions in Table \ref{tab:1} for the bosonic \eqref{eqn:bose_hamil} and the fermionic \eqref{eqn:fermi_hamil} systems with dual contact interactions, in this section, we will study the asymptotic behavior of the work distributions in the classical limit ($\hbar\rightarrow0$ or $T\rightarrow+\infty$). We will use Lieb-Liniger model ($V_{ext}(\mathbf{x},\lambda(t))$ in Eq. \eqref{eqn:bose_hamil} is a ring potential) with the coupling strength $C$ as an example.

In Table I, for various coupling strength $C$, we list the bosonic and the fermionic systems with dual contact interactions. In the classical limit, Table \ref{tab:1} is reduced to Table \ref{tab:2}. In this limit, both free Bosons and free Fermions behave like noninteracting distinguishable particles \cite{gong2014interference} (see the second and third rows in Table \ref{tab:2}). The systems belong to the first row of Table \ref{tab:1}, in which the coupling strength $C$ is finite, behave like their classical counterparts described by the corresponding Hamiltonians. Since the duality relation is independent of the values of $\hbar$ and $T$, the relation still holds in the classical limit. Namely, the systems belong to the same row of Table~\ref{tab:2} have the same work distribution. As shown in the second and third rows of Table~\ref{tab:2}, the work distributions of free Bosons and free Fermions correspond to the work distribution of noninteracting distinguishable particles \cite{gong2014interference}. Since the work distributions of the free Bosons and free Fermions is identical to the work distributions of the limiting cases ($C\rightarrow 0$ and $C\rightarrow \infty$) of Lieb-Liniger model respectively, it is reasonable to infer that for an arbitrary coupling strength $C>0$, the work distribution still converges to that of the noninteracting distinguishable particles. Intuitively, one can understand the result from the equivalence between the noninteracting distinguishable particles and the (elastic) hard-core gas in classical mechanical picture \footnote{Consider the collision process between two particles in such one dimensional systems. Because of the conservation laws of momentum and energy, for identical particles, the reflection event is indistinguishable from the transmission event.}. Numerical results in Fig. \ref{fig:2} show the tendency of the convergence of work distributions as the initial temperature increases. In the following we give a quantitative analysis to justify the convergence of work distributions of Lieb-Liniger model \cite{lieb1963exact,lieb1963exact2} with various coupling strengths $C$ in the classical limit.

As an illustration, let us consider the work distributions of Lieb-Liniger model in the quantum adiabatic process (infinitely slow change of the work parameter $\lambda(t)$). The coupling strength $C$ can be an arbitrary number. For convenience, suppose the system is in a ring with the circumference $\lambda$ and the periodic boundary condition is assumed. We set $2m=1$. When $\lambda$ is changed from $\lambda_i$ to $\lambda_f$, work is performed to the system. By definition \eqref{eqn:work_dist} and the adiabatic theorem, there is no interstate transition. i.e. $P(n_f|n_i)=\delta_{n_i,n_f}$. Thus, the work distribution is reduced to
\begin{equation}
\label{eqn: adiabatic_work}
P(W)=\sum_{n_i} P_i(n_i) \delta(W-(E_{n_i}(\lambda_f)-E_{n_i}(\lambda_i))).
\end{equation}
Our goal is to estimate Eq. \eqref{eqn: adiabatic_work} for Lieb-Liniger model with an arbitrary $C$ in the classical limit $T\rightarrow +\infty$ ($\beta\rightarrow 0$) or $\hbar \rightarrow 0$.

The eigenenergies $E_{n}(\lambda)$ are determined through $E_n(\lambda) \equiv E_{I_1,I_2...I_N}(\lambda)\equiv E_{\bf{I}}(\lambda)=\sum_{i}\hbar^2k_{i}(\lambda)^2$, where $\left \{I_i \right \}_{i=1}^{N}$ is a complete set of quantum numbers for an energy eigenstate. Here $\left \{k_i\right \}_{i=1}^N$ are solutions to the transcendental equations \cite{lieb1963exact}:
\begin{eqnarray}
\begin{aligned}
\label{eqn: transcendental}
k_l\lambda &= 2\pi I_l-2\sum_{j=1}^{N}\theta(k_l-k_j), \quad l=1,2,...N\\
& I_l \in \mathbb{N}, \qquad \text{if} \quad N=\text{odd}, \\
& I_l \in \mathbb{N}+\frac{1}{2}, \qquad \text{if} \quad N=\text{even},
\end{aligned}
\end{eqnarray}
where $\theta(k)=\text{arctan}\frac{2\hbar^2k}{C}$.

It has been proved \cite{yang1969} that for a given set of $\left \{I_i\right \}_{i=1}^{N}$ ($I_1<I_2...<I_N$), Eq \eqref{eqn: transcendental} has a unique set of solutions for $\left \{k_i\right \}_{i=1}^{N}$ ($k_1<k_2<...k_N$), which determines a unique energy eigenstate of the system.

We introduce the characteristic function of Eq. \eqref{eqn: adiabatic_work} as follows:
\begin{eqnarray}
\label{characteristics}
\begin{aligned}
&G(\nu)=\int^{+\infty}_{-\infty}P(W)e^{i\nu W} \\
&=
\frac{\sum_{\bf{I}}\text{exp}(-\beta E_{\bf{I}}(\lambda_i))\text{exp}(i\nu(E_{\bf{I}}(\lambda_f)-E_{\bf{I}}(\lambda_i)))}{\sum_{\bf{I}}\text{exp}(-\beta E_{\bf{I}}(\lambda_i))},
\end{aligned}
\end{eqnarray}
where the summation is over all possible $I_1<I_2...<I_N$ in Eq \eqref{eqn: transcendental}.

The $n$th moment of the work distribution is
\begin{equation}\label{eqn: moments}
\left \langle W^n  \right \rangle=Z^{-1}\sum_{\bf{I}}\exp(-\beta E_{\bf{I}}(\lambda_i))(E_{\bf{I}}(\lambda_f)-E_{\bf{I}}(\lambda_i))^n.
\end{equation}

In the classical limit $\beta \rightarrow 0$ or $\hbar \rightarrow 0$, exp$(-\beta E_{\bf{I}}(\lambda_i))$ $\sim$ 1. So in this limit Eq. \eqref{eqn: moments} is dominated by the terms with large values of work. Let's consider the work related to an energy eigenstate characterized by the quantum numbers $\bf{I}$$=(I_1,I_2..I_N)$,
\begin{eqnarray}
\label{eqn: energy_diff}
\begin{aligned}
&E_{\bf{I}}(\lambda_f)-E_{\bf{I}}(\lambda_i)=\hbar^2\sum_{l=1}^{N}(k_l(\lambda_f)^2-k_l(\lambda_i)^2)\\
&=\sum_{l=1}^{N}4\hbar^2 [ (\frac{1}{\lambda_f^2}-\frac{1}{\lambda_i^2}) \pi^2 I_l^2  \\
&-2\pi I_l\sum_{j=1}^{N}(\frac{\theta(k_l(\lambda_f)-k_j(\lambda_f))}{\lambda_f^2}\\
&-\frac{\theta(k_l(\lambda_i)-k_j(\lambda_i))}{\lambda_i^2}) \\
&+\frac{\sum_{j,s=1}^{N}\theta(k_l(\lambda_f)-k_j(\lambda_f))\theta(k_l(\lambda_f)-k_s(\lambda_f))}{\lambda_f^2}\\
&-\frac{\sum_{j,s=1}^{N}\theta(k_l(\lambda_i)-k_j(\lambda_i))\theta(k_l(\lambda_i)-k_s(\lambda_i))}{\lambda_i^2}].
\end{aligned}
\end{eqnarray}

The first term on the right hand side of Eq. \eqref{eqn: energy_diff} is much larger than the remaining terms when $|I_l|\gg \pi N/2$ for $l=1,2... N$, because $|\sum_{j=1}^N \theta(k)|<N/2$.

As a result, when $|I_l| \gg \frac{N}{2}$ for $l=1,2... N$, we have
\begin{eqnarray}
\begin{aligned}
E_{\bf{I}}(\lambda_f)-E_{\bf{I}}(\lambda_i)&\simeq \sum_{l=1}^{N} 4\hbar^2(\frac{1}{\lambda_f^2}-\frac{1}{\lambda_i^2}) \pi^2 I_l^2,
\end{aligned}
\end{eqnarray}
which is the eigenenergy difference of $N$ noninteracting distinguishable particles in a ring. Note that in the classical limit, the dominate contributions in Eq. \eqref{eqn: moments} come from those states in which $|I_l|\gg N/2$. Therefore, from the above analysis, we find that in the classical limit $\beta \rightarrow 0$ or $\hbar \rightarrow 0$ the $n$th moment of work distribution is approximately independent of $C$ and is approximately equal to that of noninteracting distinguishable particles (Boltzamann particles). Alternatively,
\begin{equation}
\label{moment}
\left \langle W^n \right \rangle \simeq \left \langle W^n \right \rangle_{cl}, \quad \beta \rightarrow 0 \quad \text{or} \quad \hbar \rightarrow 0,
\end{equation}
where $\left \langle W^n \right \rangle_{cl}$ denotes the $n$-th moment of the work distribution for noninteracting distinguishable particles in a ring.

Eq. \eqref{moment} tells us that for an arbitrary $C$ all the moments of the work distributions of the Lieb-Liniger model are (asymptotically) equal to those of noninteracting distinguishable particles in a ring in the classical limit. Based on the following relation,
\begin{equation}
G(\nu)=1+\sum_{n=1}^{+\infty}\frac{(i\nu)^{n}}{n!}\left \langle W^n \right \rangle,
\end{equation}
it is straightforward to show that in the classical limit the characteristic function of the work distributions of the Lieb-Liniger models with various coupling strengths converges
\begin{equation}
\label{distribution}
G(\nu)\simeq G(\nu)_{cl}, \quad \beta \rightarrow 0 \quad \text{or} \quad \hbar \rightarrow 0,
\end{equation}
where $G(\nu)_{cl}$ is the characteristic function of work distribution for $N$ noninteracting distinguishable particles in a ring. Please note that in Ref. \cite{gong2014interference}, $G(\nu)_B\simeq G(\nu)_F \simeq G(\nu)_{cl}$ in the classical limit. Since $C$ of the Lieb-Liniger model can be an arbitrary number in Eq. \eqref{distribution}, $G(\nu)_B$ an $G(\nu)_F$ in Ref. \cite{gong2014interference} correspond to two limiting cases ($C\rightarrow 0$ and $C\rightarrow \infty$) of our result. We would like to emphasize that Eq.\eqref{distribution} is the second main result of our paper \footnote{Eq. \eqref{distribution} is derived for the quantum adiabatic process, but it is also valid for an arbitrary nonequilibrium process.}.

The systems in the same column of Table \ref{tab:1} lead to different quantum work distributions when driven under the same protocol, which is due to different quantum statistics obeyed by the constituent (quasi-) particles of the systems, namely the Bose-Einstein, the Fermi-Dirac or the fractional statistics \cite{haldane1991,wu1994,shankar1994,isakov1994statistical}. But in the classical limit, such differences vanish. Namely, no matter what quantum statistics they obey, in the classical limit, their work distributions converge to the work distribution of noninteracting distinguishable particles which satisfy Boltzmann statistics. A relevant result is that in the classical limit, for an arbitrary coupling strength $C$, the equation of state of the Lieb-Liniger model \eqref{eqn:bose_hamil} converge to that of the noninteracting distinguishable particles (See Appendix \ref{Appendix:C}). Our result can be regarded as the dynamical extension of the results in  Ref. \cite{tonks1936complete}, where it was found that noninteracting distinguishable particles and classical impenetrable particles have the same equation of state.

\section{\label{sec:5}Discussion and Conclusion}
An intuitive explanation for the duality of the work distributions can be given as follows. As we know, interference of identical particles will influence their probability in real space. For free Bosons, symmetric permutation relation results in an ``effective attractive interaction". For free Fermions, antisymmetric permutation relation results in an ``effective repulsive interaction". In our model, a repulsive interaction is introduced to Bosons (\ref{eqn:bose_hamil}) and an attractive one to fermions (\ref{eqn:fermi_hamil}) to partially cancel the ``effective" interactions. Bose-Fermi duality is cancelled out by the duality of the interactions. Some properties of the two systems become identical. In particular, work distributions of Bosons \eqref{eqn:bose_hamil} and Fermions \eqref{eqn:fermi_hamil} with dual contact interactions are identical.

One of the applications of the duality \eqref{eqn:work_distrition_duality} is to use it as a bridge for calculating work distributions. The work distribution of one system can be obtained by calculating the work distribution in of its dual system. For example, it is hard to obtain the work distributions of the TG and FTG gases directly because both of them are strongly interacting quantum many-body systems. Through the dynamical Bose-Fermi duality, the problem can be reduced to the calculation of the work distribution of noninteracting Fermions and Bosons, which is obviously much simpler than the original one.

In summary, in this article, we extend the well-known static duality \cite{girardeau1960relationship, cheon1999fermion} between 1-D Bosons and 1-D Fermions to the dynamical version (\ref{eqn:dyanmical-duality}). By utilizing this dynamical duality we find the Bose-Fermi duality of quantum work distributions between 1-D Bosons \eqref{eqn:bose_hamil} and Fermions \eqref{eqn:fermi_hamil} with dual contact interaction. Particularly, we find TG gas and 1-D free fermions, though differing significantly from each other in momentum distributions \cite{lenard1964,olshanii1998atomic}, have identical quantum work distributions. In the classical limit ($\beta \rightarrow 0$ or $\hbar \rightarrow 0$), we find that work distribution of the Lieb-Liniger model (Cheon-Shigehara model) with an arbitrary coupling strength, converges to that of the noninteracting distinguishable particles. These results bring important insights to the understanding of the effects of the interplay between quantum statistics and interactions on quantum work in interacting quantum many-body systems.
\begin{acknowledgments}
H. T. Quan gratefully acknowledges support from
the National Science Foundation of China under grants
11775001, 11534002, and The Recruitment Program of
Global Youth Experts of China. Jingning Zhang gratefully acknowledges the support from the National Natural Science Foundation of China (Grants No. 11504197).
\end{acknowledgments}

\appendix

\section{Proof of the Dyanmical Bose-Fermi Duality \eqref{eqn:dyanmical-duality}}
\label{Appendix: A}
In this appendix, we prove the dynamical Bose-Fermi duality, which asserts that if the wave function of the initial states of the two systems are related by Eq.~(\ref{eqn:3}), then the time evolution of the wave functions is always related by  Eq.~(\ref{eqn:dyanmical-duality}).

Let us consider the time-independent Schr$\ddot{\rm o}$dinger equation \eqref{eqn:time_independent_schr}. Note that the contact interactions for both the bosonic \eqref{eqn:bose_hamil} and fermionc \eqref{eqn:fermi_hamil} systems have nonvanishing effects only when the coordinates of two particles overlap. As a consequence, in the region $R_1(t)\subset \mathbb{R}^N$ with $0\leq x_1\leq x_2\leq\ldots\leq x_N\leq\lambda(t)$, the Hamiltonians of both the interacting Bosons \eqref{eqn:bose_hamil} and the interacting Fermions \eqref{eqn:fermi_hamil} coincide with that of the noninteracting particles,
\begin{eqnarray*}
\hat H_{R_1}^{B,F}(\lambda(t))=\hat H_0(\lambda(t))=\sum_{i=1}^N\frac{\hat p_i^2}{2m}+\sum_{i=1}^N V_{\rm ext}(\hat x_i,\lambda(t)).
\end{eqnarray*}
Thus in region $R_1$, the time-independent Schr$\ddot{\rm o}$dinger equation \eqref{eqn:time_independent_schr} is reduced to the following equation,
\begin{eqnarray}
\label{eqn:schrodinger_R1}
\hat H_0(\lambda(t))\phi_n^{B, F}\left({\mathbf x};\lambda(t)\right)=E^{B, F}_n(\lambda(t))\phi_n^{B,F}\left({\mathbf x};\lambda(t)\right),
\end{eqnarray}
with the absorbing boundary condition
\begin{eqnarray}
\label{eqn:absorb_bc}
\left.\phi_n^{B, F}({\mathbf x};\lambda(t))\right|_{x_i=0 or \lambda(t)}=0.
\end{eqnarray}

Except for the physical boundaries $x_i=0$ or $\lambda(t)$ induced by the external potential, there are additional boundaries determined by $x_i = x_j$ with $i\neq j$, which physically means that the particle $i$ and particle $j$ overlap. At this boundary, the wave functions for bosonic \eqref{eqn:bose_hamil} and fermionic \eqref{eqn:fermi_hamil} systems connect in different ways.

For the bosonic system, the energy eigenstates satisfy the following relations at the boundary $x_i=x_j$,
\begin{eqnarray}
\label{eqn:bose_collision_boundary}
\left.\phi_n^{B}({\mathbf x};\lambda(t))\right|_{x_i=x_j^-}&=&\left.\phi_n^{B}({\mathbf x};\lambda(t))\right|_{x_i=x_j^+},\\
 \frac{\partial \phi_n^B({\mathbf x};\lambda(t))}{\partial x_i}|_{x_i=x_j^-}&=&-\frac{\partial \phi_n^B({\mathbf x};\lambda(t))}{\partial x_i}|_{x_i=x_j^+},\nonumber
\end{eqnarray}
while for the fermionic system, the energy eigenstates satisfy the following relations,
\begin{eqnarray}
\label{eqn:fermi_collision_boundary}
\left. \phi_n^{F}({\mathbf x};\lambda(t))\right|_{x_i=x_j^-}&=&-\left.\phi_n^{F}({\mathbf x};\lambda(t))\right|_{x_i=x_j^+},\\
\frac{\partial \phi_n^F({\mathbf x};\lambda(t))}{\partial x_i}|_{x_i=x_j^-}&=&\frac{\partial \phi_n^F({\mathbf x};\lambda(t))}{\partial x_i}|_{x_i=x_j^+}.\nonumber
\end{eqnarray}
The physical intuition of the above two sets of boundary conditions is that at the boundaries, the bosonic energy eigenstates are continuous while their spatial derivatives are reversed, and the situation is interchanged for the fermionic energy eigenstates.

Then we turn to investigate effects of the contact interactions on the bosonic and fermionic energy eigenstate. Note that the contact interactions have non-vanishing effects only when two particles overlap. The effect of the $\delta$-function potential in the bosonic system with respect to $x_i$ and $x_{i+1}$ can be expressed as follows,
\begin{eqnarray}
\label{eqn:bose_potential}
\frac{\partial\phi_n^B({\mathbf x};\lambda(t))}{\partial x_i}|^{x_{i+1}^+}_{x_{i+1}^-}&=&\left.\frac{mC}{\hbar^2}\phi_n^B({\mathbf x};\lambda(t))\right|_{x_i=x_{i+1}},\\
\frac{\partial \phi_n^B({\mathbf x};\lambda(t))}{\partial x_{i+1}}|^{x_i^+}_{x_i^-}&=&\left.\frac{mC}{\hbar^2}\phi_n^B({\mathbf x};\lambda(t))\right|_{x_{i+1}=x_i}.\nonumber
\end{eqnarray}
While for the fermionic system, the $\varepsilon$-function potential relates the energy eigenstates and their spatial derivatives at the boundary in the following way~\cite{cheon1998realizing},
\begin{eqnarray}
\label{eqn:fermi_potential}
\left.\phi_n^F({\mathbf x};\lambda(t))\right|^{x_i=x_{i+1}^+}_{x_i=x_{i+1}^-}&=&\frac{\hbar^2}{mC}\frac{\partial\phi_n^F({\mathbf x};\lambda(t))}{\partial x_i}|_{x_{i+1}^-},\\
\left.\phi_n^F({\mathbf x};\lambda(t))\right|^{x_{i+1}=x_i^+}_{x_{i+1}=x_i^-}&=&\frac{\hbar^2}{mC}\frac{\partial\phi_n^F({\mathbf x};\lambda(t))}{\partial x_i}|_{x_i^-}.\nonumber
\end{eqnarray}

Combining Eqs.~(\ref{eqn:bose_collision_boundary}) and (\ref{eqn:bose_potential}) for the bosonic system,  and Eqs.~(\ref{eqn:fermi_collision_boundary}) and (\ref{eqn:fermi_potential}) for the fermionic system, we obtain the extra boundary condition for energy eigenstates in region $R_1$ as follows,
\begin{eqnarray}
\begin{aligned}
\label{eqn:extra_bc_R1}
-(\frac{\partial}{\partial x_i}+\frac{\partial}{\partial x_{i+1}})\phi_n^{B, F}({\mathbf x};\lambda(t))|_{x_i=x_{i+1}^-}\\=\left.\frac{mC}{\hbar^2}\phi_n^{B, F}({\mathbf x};\lambda(t))\right|_{x_i=x_{i+1}^-}.\\
\end{aligned}
\end{eqnarray}

It is clear at this stage that in region $R_1$, the time-independent Schr$\ddot{\rm o}$dinger equation (\ref{eqn:schrodinger_R1}), the absorbing boundary condition (\ref{eqn:absorb_bc}), and the extra boundary conditions (\ref{eqn:extra_bc_R1}) for the bosonic and fermionc systems are the same. Thus it is natural that the bosonic and fermionc systems have identical instantaneous energy spectra and their instantaneous energy eigenstates in this region are the same \cite{cheon1999fermion},
\begin{eqnarray}
E_n^B(\lambda(t))&=&E_n^F(\lambda(t)),\\
\left.\phi_n^B({\mathbf x};\lambda(t))\right|_{{\mathbf x}\in R_1}&=&\left.\phi_n^F({\mathbf x};\lambda(t))\right|_{{\mathbf x}\in R_1}.\nonumber
\end{eqnarray}

The $N$-particle instantaueous energy eigenstates defined in other regions can be obtained by the permutation operation
\begin{eqnarray*}
\phi_n^{B,F}({\mathbf x};\lambda(t))=\left(\pm 1\right)^P\phi_n^{B,F}\left(P({\mathbf x});\lambda(t)\right)|_{P(\mathbf{x})\in R_1}
\end{eqnarray*}
with the $+$ ($-$) sign for the bosonic (fermionic) system. Here $P({\mathbf x})=(x_{P_1},x_{P_2},\ldots, x_{P_N})$ is a permutation of ${\mathbf x}$ that satisfies $0\leq x_{P_1}\leq x_{P_2}\leq\ldots\leq x_{P_N}\leq\lambda(t)$ with $P$ denoting the permutation operation. Thus it's straightforward to see that there is a one-to-one mapping between the instantaneous energy eigenstates for the bosonic \eqref{eqn:bose_hamil} and fermionic \eqref{eqn:fermi_hamil} systems \cite{cheon1999fermion},
\begin{eqnarray}
\label{eqn:instan_eigenstate}
\phi_n^F({\mathbf x};t)=A({\mathbf x})\phi_n^B({\mathbf x};\lambda(t)),
\end{eqnarray}
where $A({\mathbf x})=\prod_{i>j}{\rm sign}(x_i-x_j)$.

In order to reveal the dyanmical Bose-Fermi duality, we expand the time-dependent wave functions $\psi^{B, F}({\mathbf x}, t)$ in terms of the instantaneous energy eigenstates,
\begin{eqnarray}
\label{eqn:expansion}
\psi^{B, F}({\mathbf x}, \lambda(t)) = \sum_{n} C_n^{B, F}(t) \phi_n^{B, F}({\mathbf x}; \lambda(t)).
\end{eqnarray}
The time-dependent Schr$\ddot{\rm o}$dinger equation \eqref{eqn:schrodinger} can be written as a set of differential equations for the expansion coefficients,
\begin{eqnarray}
i\hbar\dot C_n^{B, F}(t)&=&E_n(\lambda(t))C_n^{B, F}(t)\\
&&-i\hbar\sum_mC_m^{B, F}(t)\overlap{\phi_n^{B, F}(\lambda(t))}{\partial_t\phi_m^{B, F}(\lambda(t))},\nonumber
\end{eqnarray}
with $\overlap{\mathbf x}{\phi_n^{B, F}(\lambda(t))}\equiv\phi_n^{B, F}({\mathbf x}; \lambda(t))$. By using the fact $\left|A({\mathbf x})\right|^2=1$, it is easy to find that
\begin{eqnarray*}
\overlap{\phi_n^B(\lambda(t))}{\partial_t\phi_m^B(\lambda(t))}=\overlap{\phi_n^F(\lambda(t))}{\partial_t\phi_m^F(\lambda(t))}.
\end{eqnarray*}

Thus the time-dependent expansion coefficients $\left\{C_n^{B, F}(t)\right\}$ for the bosonic \eqref{eqn:bose_hamil} and fermionic \eqref{eqn:fermi_hamil} systems obey the same set of differential equations. With the same initial conditions
\begin{eqnarray}
C_n^B(0) = C_n^F(0),
\end{eqnarray}
which is equivalent to the initial condition for the wave functions \eqref{eqn:3}, one can obtain the following set of relations,
\begin{eqnarray}
\label{eqn:coefficients}
C_n^B(t)=C_n^F(t).
\end{eqnarray}

Combining Eqs.~(\ref{eqn:instan_eigenstate}), (\ref{eqn:expansion}) and (\ref{eqn:coefficients}), we prove the dynamical Bose-Fermi duality \eqref{eqn:dyanmical-duality}.

\section{Equation of State in the Classical Limit}
\label{Appendix:C}
To see the quantum-classical transition more clearly, we now consider the equation of state for the Lieb-Liniger model and calculate the quantum corrections to the equation of state of the noninteracting distinguishable particles. According to Ref. \cite{yang1969}, in the thermodynamic limit, the equation of state of Lieb-Liniger model is determined by $\epsilon(k)$. $\epsilon(k)$ satisfies the following integral equation
\begin{eqnarray}
\label{eqn:integralequation}
\begin{aligned}
\epsilon(k)=-\mu+\hbar^2k^2-\frac{2C}{\pi\beta}&\int_{-\infty}^{+\infty}\frac{\hbar^3dq}{C^2+\hbar^4(k-q)^2}
\\
&\times \ln(1+\text{exp}(-\beta\epsilon(k))),
\end{aligned}
\end{eqnarray}
where $\mu$ is the chemical potential of the system.

It has been shown in Ref. \cite{yang1969} that the solution of Eq. \eqref{eqn:integralequation}, $\epsilon(k,\beta,\mu)$ is analytical in the neighborhood of any real pair $(\beta,\mu)$. So we can expand exp$(\beta\epsilon(k))$ in terms of the fugacity $z=$exp$(\beta \mu)$ as follows:
\begin{equation}
\label{eqn:fugacity_expand}
\exp(\beta\epsilon(k))=\sum_{n=0}^{+\infty}a_n(k,\beta)z^n.
\end{equation}

Inserting Eq. \eqref{eqn:fugacity_expand} in Eq. \eqref{eqn:integralequation}, we obtain
\begin{eqnarray}
\label{coefficients}
\begin{aligned}
&a_0=0,\\
&a_1=\exp(-\beta \hbar^2 k^2),\\
&a_2=-\frac{2C}{\pi}\int_{-\infty}^{+\infty}\frac{\hbar^3e^{-\beta\hbar^2k^2}dq}{c^2+\hbar^4(k-q)^2}.\\
\end{aligned}
\end{eqnarray}

According to Ref. \cite{yang1969} the pressure can be written as
\begin{eqnarray}
\label{pressure}
P&=\frac{1}{2\pi\beta}\int_{-\infty}^{+\infty}\ln(1+e^{-\beta\epsilon(k)})dk\\
 &=\frac{1}{2\pi\beta}(b_1z+b_2z^2+O(z^3)),
\end{eqnarray}
where $b_1=\int_{-\infty}^{+\infty}a_1dk=2\pi(\sqrt{\beta}\hbar)^{-1}$, $b_2=\int_{-\infty}^{+\infty}(a_2-a_1^2/2)dk$.

The density of particle number is given by $D=\partial P/\partial A=\beta z\partial P/\partial z$, which can be written as
\begin{equation}
\label{density}
D=\frac{1}{2\pi}(b_1z+2b_2z^2+O(z^3)).
\end{equation}

Combining Eqs. \eqref{coefficients}, \eqref{pressure} and \eqref{density}, we obtain the Virial expansion (see Eq. (E3.9) of Ref. \cite{vsamaj2013introduction}) of the equation of state
\begin{equation}
\label{eqn: equationofstate}
\frac{P\beta}{D}=1-b_2\sqrt{\beta}D+O(D^2).
\end{equation}
It's not hard to find that when $\beta\hbar^2 \rightarrow 0$, we have $b_2 \rightarrow 0$ and the quantum corrections vanish. Eq. \eqref{eqn: equationofstate} becomes $P=Dk_BT$, which is the well-known equation of state for the noninteracting distinguishable particles.
\bibliography{duality}

\end{document}